\documentclass[conference]{IEEEtran}

\IEEEoverridecommandlockouts

\usepackage[absolute,showboxes]{textpos}
\TPMargin{5pt}
\newcommand{\copyrightstatement}{
\begin{textblock}{0.84}(0.08,0.93)    
\noindent
\footnotesize
\copyright 2020 IEEE. Personal use of this material is permitted. Permission from IEEE must be obtained for all other uses, in any current or future media, including reprinting/republishing this material for advertising or promotional purposes, creating new collective works, for resale or redistribution to servers or lists, or reuse of any copyrighted component of this work in other works. https://doi.org/10.1109/IJCNN48605.2020.9207719
\end{textblock}
}

\usepackage{cite}
\usepackage{amsmath,amssymb,amsfonts}
\usepackage{algorithmic}
\usepackage{graphicx}
\usepackage{textcomp}
\usepackage{xcolor}
\usepackage{booktabs}
\usepackage{threeparttable}

\usepackage[normalem]{ulem}
\usepackage{verbatim}

\renewcommand\sout{\bgroup\markoverwith
{\textcolor{red}{\rule[.5ex]{2pt}{0.4pt}}}\ULon}

\def\BibTeX{{\rm B\kern-.05em{\sc i\kern-.025em b}\kern-.08em
    T\kern-.1667em\lower.7ex\hbox{E}\kern-.125emX}}
    
\begin{document}

\title{RED: Deep Recurrent Neural Networks for Sleep EEG Event Detection
\thanks{Nicol\'as Tapia acknowledges financial support from the National Agency for Research and Development (ANID) /  Scholarship Program / MAGISTER NACIONAL/2019 - 22191803. The authors acknowledge financial support from ANID-Chile through grant FONDECYT 1171678. Additionally, the authors acknowledge financial support from the Department of Electrical Engineering at Universidad de Chile.}
}

\author{
\IEEEauthorblockN{Nicol\'as I. Tapia}
\IEEEauthorblockA{
Department of Electrical Engineering\\
Universidad de Chile\\
Santiago, Chile \\
nicolas.tapia@ug.uchile.cl}
\and
\IEEEauthorblockN{Pablo A. Est\'evez}
\IEEEauthorblockA{
Department of Electrical Engineering\\
Universidad de Chile\\
Santiago, Chile \\
pestevez@cec.uchile.cl}
}

\maketitle

\copyrightstatement

\begin{abstract}
The brain electrical activity presents several short events during sleep that can be observed as distinctive micro-structures in the electroencephalogram (EEG), such as sleep spindles and K-complexes. These events have been associated with biological processes and neurological disorders, making them a research topic in sleep medicine. However, manual detection limits their study because it is time-consuming and affected by significant inter-expert variability, motivating automatic approaches. We propose a deep learning approach based on convolutional and recurrent neural networks for sleep EEG event detection called Recurrent Event Detector (RED). RED uses one of two input representations: a) the time-domain EEG signal, or b) a complex spectrogram of the signal obtained with the Continuous Wavelet Transform (CWT). Unlike previous approaches, a fixed time window is avoided and temporal context is integrated to better emulate the visual criteria of experts. When evaluated on the MASS dataset, our detectors outperform the state of the art in both sleep spindle and K-complex detection with a mean F1-score of at least 80.9\% and 82.6\%, respectively. Although the CWT-domain model obtained a similar performance than its time-domain counterpart, the former allows in principle a more interpretable input representation due to the use of a spectrogram. The proposed approach is event-agnostic and can be used directly to detect other types of sleep events.
\end{abstract}


\section{Introduction}

Sleep research plays a key role in the study of healthy brain functioning and development. Typically, a set of physiological signals called polysomnogram (PSG) is recorded, among which the electroencephalogram (EEG), i.e., the measurement of the brain electrical activity, stands out. The PSG allows experts to recognize sleep stages and short events caused by biological processes. According to the official manual of the American Academy of Sleep Medicine (AASM), sleep is divided into five stages called W, R, N1, N2 and N3 \cite{berry2012aasm}, where W, R and N stand for Wake, REM and Non-REM, respectively. 

\begin{figure}[tbp]
\centering
\includegraphics[width=0.85\columnwidth, trim={0.12in 0.17in 0.16in 0.15in}, clip]{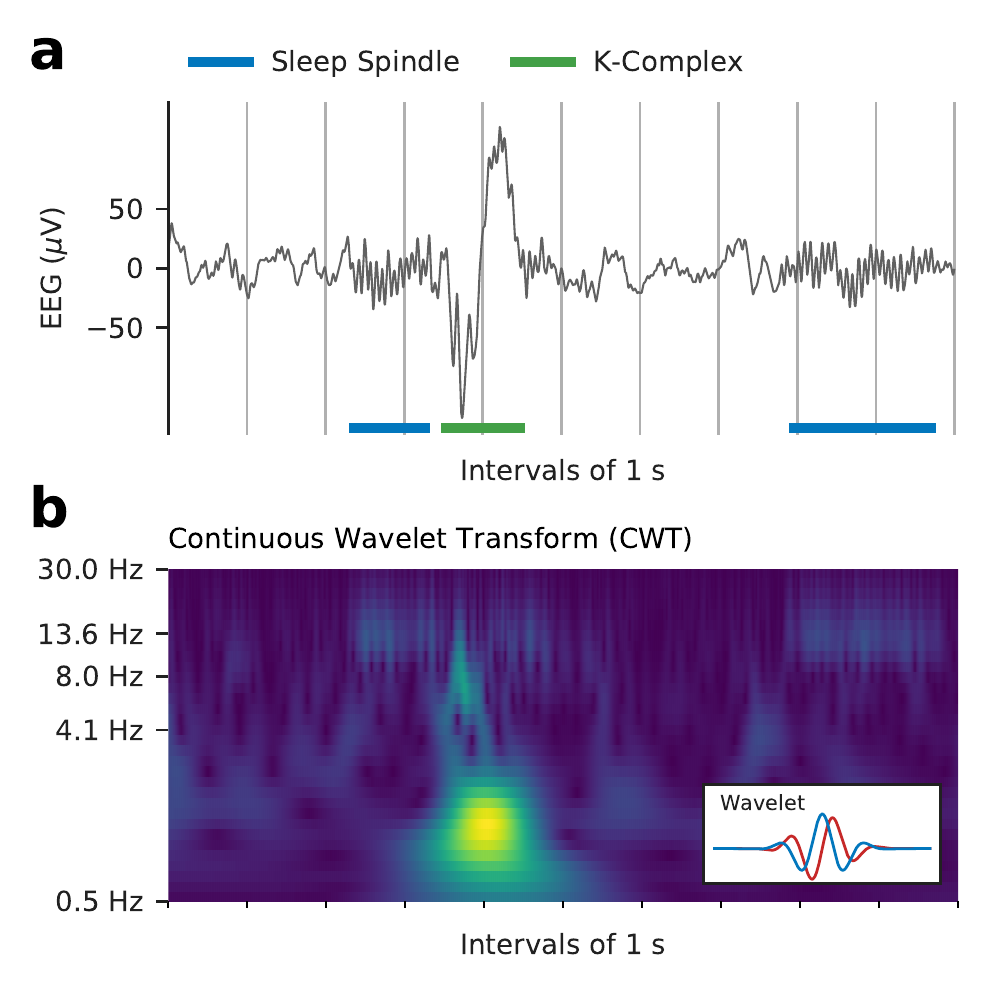}
\caption{An EEG segment during stage N2. (\textbf{a}) One K-complex and two sleep spindles annotated by a human expert. (\textbf{b}) Magnitude of the Continuous Wavelet Transform (CWT) of the same EEG segment shown in (a), obtained using the complex Morlet wavelet (shown in inset) with a geometrical progression of 32 scales between central frequencies 0.5~Hz and 30~Hz.}
\label{fig:ss_kc_cwt_example}
\end{figure}

Half of sleep is spent in stage N2, which is dominated by low-amplitude background activity of 4-7~Hz and by two landmark events called sleep spindles and K-complexes. Fig.~\ref{fig:ss_kc_cwt_example}a shows an EEG segment in stage N2 where both types of events can be observed. Sleep spindles are bursts of oscillatory activity, typically of 0.5-2~s and 11-16~Hz (sigma band) in adults \cite{berry2012aasm}, which have been associated with learning \cite{coppieters2016sleep}, memory consolidation \cite{coppieters2016sleep}, and schizophrenia \cite{clawson2016form}. K-complexes are high-amplitude biphasic waves, typically of 0.5-1.5~s and 0.5-2~Hz \cite{berry2012aasm}, which have been associated with internal and external stimulus processing \cite{wauquier1995k}, sleep maintenance \cite{wauquier1995k}, epilepsy \cite{el2008k}, obstructive sleep apnea \cite{guilleminault1976sleep}, and restless leg syndrome \cite{glasauer2001restless}.

The automatic detection of these EEG events is desirable for the following reasons. Manual detection is time-consuming because experts visually inspect the PSG in segments of 20~s or 30~s, called epochs. Furthermore, there is low inter-expert agreement because of subjective definitions \cite{warby2014sleep}. Conversely, automatic algorithms offer consistent and fast detections. However, most algorithms show a high false-discovery rate \cite{coppieters2016sleep}. Recently, deep learning methods \cite{chambon2019dosed, kulkarni2019deep} have been applied, achieving state-of-the-art performances. 

The AASM manual describes sleep spindles and K-complexes as salient patterns in the EEG, implying a concept of context. Additionally, the long-term temporal structure provides useful information for detection, such as compatible surrounding activity and signal artifacts. 


Our hypothesis is that modeling long-term temporal context can improve detection performance. In this paper we propose a deep learning approach called Recurrent Event Detector (RED) based on convolutional layers for local feature extraction and recurrent layers for long-term temporal modeling. We propose two variants of RED with different input representations: RED-Time, that uses the 1D time-domain EEG signal, and RED-CWT, that uses a 2D spectrogram obtained with the Continuous Wavelet Transform (CWT) \cite{addison2017illustrated}. 

The main contributions of this work are the following: a) to propose a detection method with better context modeling and without arbitrary input partitions, b) to compare the performance of the time-domain and the CWT-domain input representations, and c) to assess the proposed models and the baselines using a common evaluation framework.

\section{Related Work}
\label{sec:related}

Several methods have been proposed for sleep spindle and K-complex detection. They use a variety of processing techniques, either only in the time-domain \cite{chambon2019dosed, parekh2017multichannel, larocco2018spindler, held2004dual} or in combination with the frequency-domain \cite{lacourse2019sleep, kulkarni2019deep, lachner2018single, lajnef2017meet, estevez2007sleep, causa2010automated, ulloa2016sleep}. Traditional methods assign a label by sliding a window of fixed width or by thresholding one or more time-varying features. The former approach is limited by the window width because a large window allows more information to be used but decreases the temporal resolution; and the latter approach depends strongly on the designed features and the generalization capability of the thresholds to different subjects. In addition, traditional methods usually produce a high number of false positives due to the biological variability inter- and intra-subjects as well as a subjective human scoring that makes the knowledge transfer from sleep experts to an algorithm a difficult task. Moreover, hand-crafted features are event-specific so they do not generalize to other types of events. As an alternative, deep learning methods can automatically learn features from data that can generalize to unseen recordings. 

DOSED \cite{chambon2019dosed} is an event-agnostic approach based on a convolutional neural network that processes the time-domain EEG signal in segments of 20~s. After the convolutional feature extraction, the prediction is done using a single convolutional layer with valid padding whose kernel size is equal to the previous feature map size. This results in a fully-connected layer applied to the entire EEG segment to take into account the temporal context of events. Default centers with a default duration are evenly placed on the segment, and DOSED predicts whether there is an event at each location or not and the corresponding deviations from the default values.

DOSED has two main limitations. The first one is that the default duration and the spacing between default centers are two hyperparameters that need to be adjusted considering domain knowledge for the event of interest. The second one is a low capacity to model temporal context. The whole segment is combined only at the last layer, implying that long-term temporal dependencies are restricted to be linear. Additionally, even though the theoretical receptive field is large at the last convolutional layer, the effective receptive field could be much smaller \cite{luo2016understanding}, restricting how much of the local context is used.

Another deep learning method is SpindleNet \cite{kulkarni2019deep}, which is specific to sleep spindles. First, a window of 250~ms is extracted from the EEG and band-pass filtered in the sigma band. Two neural networks, each of them composed of convolutional and recurrent layers, process the original signal and the envelope of the filtered signal, respectively. The outputs are concatenated with power features computed on the same window to predict whether it belongs to a sleep spindle or not using fully-connected layers. By sliding the window through the signal, predictions are made in the entire EEG. 

The main limitation of SpindleNet is that it cannot model temporal context outside the small window of EEG. In addition, it is event-specific, and the right window width has to be selected, which affects both the temporal resolution and the maximum context used by the model. Moreover, the sliding-window approach is inefficient for offline detection due to redundant computations.

\section{Recurrent Event Detector}

In this section, we describe our proposed model called the Recurrent Event Detector (RED), a deep learning approach that does not suffer from the limitations described above for DOSED and SpindleNet. First, we describe the EEG preprocessing and the CWT \cite{addison2017illustrated} that is used to transform the 1D time-domain signal to a 2D time-frequency representation. Next, we introduce two variants of RED: RED-Time, that uses the time-domain signal directly, and RED-CWT, that uses the CWT spectrogram as input. Finally, we describe the training process and the postprocessing of predictions.

\subsection{Preprocessing}
\label{sec:preprocessing}

EEG signals are filtered with a 0.3-35~Hz band-pass filter following standard procedures \cite{berry2012aasm} and resampled to 200~Hz. Each signal is divided by the standard deviation computed from the entire non-testing set, and values below $-10$ or above $10$ were clipped. No artifact removal procedure was conducted.

\subsection{Continuous Wavelet Transform}
\label{sec:cwt}

An alternative input representation is obtained by transforming the time-domain signal into a complex time-frequency representation. There are several methods available to achieve this, e.g., the Short-Time Fourier Transform (STFT). In this work, we use the Continuous Wavelet Transform (CWT) \cite{addison2017illustrated} , which has optimum time-frequency resolution trade-off thanks to its adaptive window size at each central frequency, as opposed to the fixed-window approach of the STFT. In this way, higher frequencies are extracted using smaller windows, implying higher temporal resolution at the expense of lower frequency resolution. This characteristic makes the CWT suitable to capture transient events in the EEG.

The CWT at scale $s$ of an input signal $x(t)$ is computed by convolving $x(t)$ with a scaled wavelet $\psi_s(t)$ as
\begin{equation}
    \mathrm{CWT}[x](s,t) = \int x(\tau) \psi_s^* (t-\tau)d\tau .
    \label{eq:cwt_int}
\end{equation}
For simplicity, we use the complex Morlet wavelet to capture both amplitude and phase:
\begin{equation}
    \psi_s(t) = \frac{1}{Z(s, \beta)} \exp\left(j\frac{2\pi t}{s}\right)\exp\left(-\frac{t^2}{\beta s^2}\right).
    \label{eq:cwt_wavelet}
\end{equation}
The wavelet width $\beta$ controls the time-frequency resolution trade-off. The larger its value, the higher the frequency resolution and the lower the temporal resolution. The normalization constant $Z(s, \beta)$ is generally chosen to have unit energy.

The central frequency from which each $\psi_s(t)$ extracts power is determined by the scale $s$, where lower frequencies are associated with larger scales. We choose a geometric progression of scales, the recommended spacing, to span $N_s$ frequencies between $f_{min}$ and $f_{max}$. Motivated by the set of frequency bands used in sleep medicine \cite{berry2012aasm}, we set $f_{min}$ and $f_{max}$ to 0.5~Hz and 30~Hz, respectively. Fig.~\ref{fig:ss_kc_cwt_example}b shows the magnitude of the CWT of the EEG segment shown in Fig.~\ref{fig:ss_kc_cwt_example}a, for $N_s=32$ and $\beta=0.5$. For visualization purposes, $Z(s,\beta)$ was chosen to have unity gain at each central frequency.

Theoretically, $\psi_s(t)$ in (\ref{eq:cwt_wavelet}) has infinite support, but a finite support is used in practice. For the function $\exp(-x^2/(2\sigma^2))$, a common heuristic is to truncate it to $[-3\sigma,\ 3\sigma]$ because 99.7\% of its mass lies within this interval. Following this heuristic, we truncate each wavelet to $[-\eta s \sqrt{4.5\beta},\ \eta s \sqrt{4.5\beta}]$, where $\eta\ge 1$ can be adjusted for a wider interval if desired.

For lower frequencies, $\psi_s(t)$ stretches significantly, resulting in pronounced border effects. To ensure an accurate representation, a padding of $T_B$ samples from the EEG is used to compute the CWT. The value of $T_B$ has to be sufficiently large to ensure that there are no border effects inside the segment of interest. After obtaining the CWT, the additional samples are dropped. In Fig.~\ref{fig:ss_kc_cwt_example}b, the CWT was obtained using 5~s of additional samples at each side of the segment, which implies $T_B=1000$ at 200~Hz sampling frequency.

\subsection{Architectures of the Proposed Models}
\label{sec:models}

\begin{figure}[tbp]
\centering 
\includegraphics[width=\columnwidth]{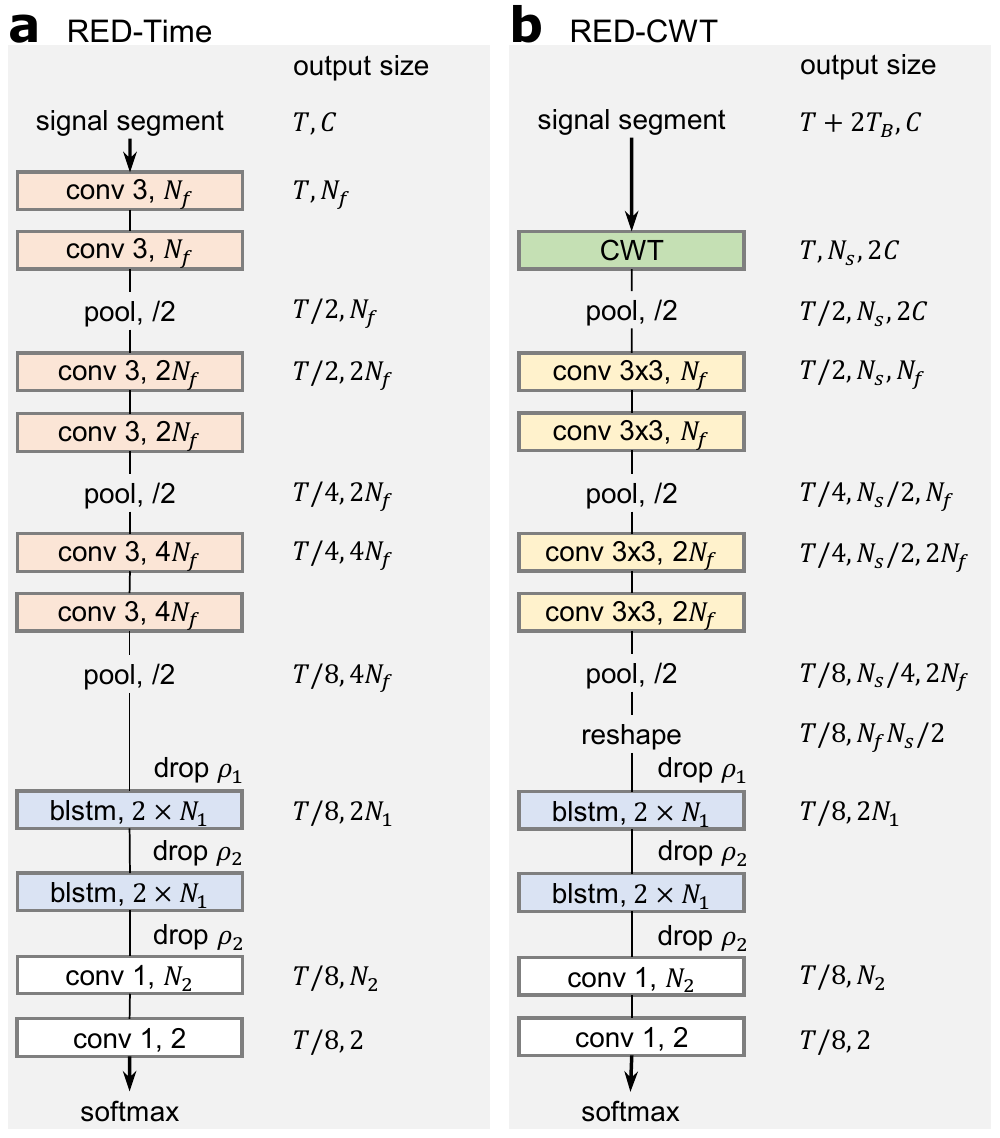}
\caption{(\textbf{a}) Time-domain model (RED-Time) and (\textbf{b}) CWT-domain model (RED-CWT). The input of RED-Time is a signal segment of $T$ samples and $C$ channels. In this work, $C=1$. For RED-CWT, a padding of $T_B$ samples is used. The output is a dense segmentation with one label every eight input samples. Batch-normalization layers are not shown. Output sizes have channel-last format.}
\label{fig:model}
\end{figure}

Fig.~\ref{fig:model} illustrates our proposed time-domain and CWT-domain models, called RED-Time and RED-CWT, respectively. The input is a preprocessed signal segment of $T$ samples and $C$ channels. A padding of $T_B$ samples is used for RED-CWT (see Sec. \ref{sec:cwt}). In this work, we study the single-channel case only ($C=1$). Additionally, we set $T=4000$ and $T_B=1000$, equivalent to 20~s and 5~s at 200~Hz sampling rate, respectively. Both models are composed of three main parts: local feature extraction, temporal context modeling, and by-sample classification.

The local feature extraction consists of a stack of convolutional layers with an initial number of filters $N_f$, a kernel of size 3, zero-padding, ReLU activation, and batch-normalization \cite{ioffe2015batch} after each convolution and before ReLU. Downsampling is performed after two convolutions using a pooling layer of size 2, and the number of filters is doubled after pooling. Batch-normalization is also applied before the first convolutional layer. This helps to randomly scale the input during training, introducing regularization. In RED-Time, the 1D convolutional layers are applied directly to the time-domain input. On the other hand, in RED-CWT the 2D convolutional layers are applied to the CWT spectrogram of the input. The CWT is computed using a complex Morlet wavelet of width $\beta$ and $N_s$ scales. The real and imaginary parts for each input channel are concatenated in the channel axis, and a pooling layer is applied only in the time axis. In this case, the first batch-normalization layer normalizes across the frequency axis, which allows the model to learn a frequency-specific normalization. For this reason, we simply set $Z(s, \beta)=1$. We let the model learn the width $\beta$. Because each wavelet size is determined at the initialization, we set $\eta=1.5$ to allow the width to increase if necessary. At the end of this stage, the output of both RED-Time and RED-CWT is a multivariate time series of length $T/8$. Following this first stage of local feature extraction, both models have the same architecture.

Temporal context modeling is achieved using Long Short-Term Memory (LSTM) layers \cite{hochreiter1997long}, a class of recurrent layers that can learn long-term dependencies thanks to a gating mechanism designed to alleviate the vanishing gradient problem of vanilla recurrent layers. Two LSTM layers are concatenated to form a Bidirectional LSTM (BLSTM) layer \cite{graves2005framewise}, which processes the input in the forward and backward directions. In this way, each hidden state has left and right contexts. We use two BLSTM layers with $N_1$ units per direction to process the time series of extracted local features. Dropout \cite{srivastava2014dropout} is applied to the input of each BLSTM layer with dropout rates $\rho_1$ and $\rho_2$, respectively. 

By-sample classification is achieved by two 1D convolutional layers with kernels of size 1. They are effectively two fully-connected layers with weights shared across time. The first layer has $N_2$ filters and ReLU activation, whereas the last layer has 2 filters and softmax activation. Dropout is applied to the input of the first layer with dropout rate $\rho_2$. The classification stage maps each time step to the probability of belonging to the background (class 0) or to an event of interest (class 1). The final output has a temporal resolution eight times lower than the original input signal. For a 200~Hz sampling rate, this means that labels are predicted every 40~ms instead of every 5~ms, which is still dense enough for events like sleep spindles and K-complexes that last at least 0.3~s.

\subsection{Training and Inference}
\label{sec:train}

The models are trained to minimize the cross-entropy loss with respect to the binary sequence of expert labels representing the existence of an event at each time step. We use the Adam optimizer \cite{kingma2014adam} with learning rate $\alpha$, default exponential decay rates, and batch size $M$. The learning rate is halved when the validation loss has not improved for $I_{lr}$ iterations, and training is stopped after $N_{lr}=4$ halving steps. For recurrent architectures, it is recommended to clip the gradient to avoid exploding gradients during training \cite{pascanu2013difficulty}. Therefore, the gradient is restricted to a maximum global norm $g_{max}=1$. 

The extraction of EEG segments at training and inference times is different. To augment the number of available segments and to improve the robustness to temporal translations, we randomly crop the recordings at training time. At each iteration, we sample $M$ epochs from the training set and we randomly choose a center inside each epoch. Finally, those random centers are used to extract the $M$ segments of size $T$. At inference time we are interested in avoiding border effects. Therefore, predictions are made in each recording using segments extracted sequentially with stride $T/2$. Then, the central $T/2$ samples of each prediction are kept.

In practice, events such as sleep spindles and K-complexes are rare compared to the background EEG activity. For this reason, we count the number of samples belonging to an event inside each epoch of 20s in the training set and compute the median value to balance the batches. At each iteration, $M/2$ examples are extracted from epochs below the median and the other $M/2$ examples from epochs above the median.

To return the final detection at inference time, the probability output of a recording is linearly upsampled by eight to achieve the original sampling rate. Then, each sample is assigned to the class 1 if the probability of class 1 exceeds an output threshold $\mu$, and to the class 0 otherwise. Finally, this binary sequence is transformed to a list of start and end time steps representing the locations of the predicted events.

\subsection{Postprocessing}
\label{sec:postprocessing}

Event-specific postprocessing is applied to the predicted events. For sleep spindle detection, following standard procedures \cite{berry2012aasm} but with a lower minimum duration as in \cite{warby2014sleep}, we combine predictions that are closer than 0.3~s and remove predictions shorter than 0.3~s. In practice, sleep spindles are rarely longer than 3~s \cite{warby2014sleep}, so we remove predictions longer than 5~s, and predictions between 3~s and 5~s are cropped to keep the central 3~s. On the other hand, K-complexes should last at least 0.5~s according to \cite{berry2012aasm}, so we remove predictions shorter than 0.3~s to allow some prediction error. Because multiple K-complexes could be predicted as a single event due to no minimum separation, we apply a splitting procedure inspired by \cite{lajnef2017meet}. Within each prediction, the low-pass filtered signal with 4~Hz cut-off is used to find negative peaks. To avoid border artifacts, we ignore peaks closer than 0.05~s to the start of the prediction or closer than 0.2~s to the end. Next, peaks without a zero-crossing in between are averaged, so that each K-complex candidate is represented by a single negative peak. If more than one peak is left after this procedure, we split the prediction by the middle point between those peaks.

\section{Experiments}
In this section, we describe the experiments used to evaluate our models and the dataset. Next, we describe the evaluation framework based on the Intersection over Union (IoU) metric that is applied to our models and selected baselines when solving the tasks of sleep spindle and K-complex detection.

\subsection{Data: The Montreal Archive of Sleep Studies}
\label{sec:data_description}

The publicly available Montreal Archive of Sleep Studies (MASS) dataset \cite{o2014montreal}, subset 2 (SS2), consists of 19 annotated whole-night PSG sampled at 256~Hz of young healthy adults between 18 and 33 years old. Each PSG corresponds to a different subject. Annotations of sleep stages followed the old R\&K manual using epochs of 20~s. Following standard practice, stage 2 of R\&K is considered to be equivalent to stage N2 of AASM. K-complex and sleep spindle annotations were provided by two experts E1 and E2 on N2 epochs of the C3-CLE EEG channel. Both events were annotated by E1 in all recordings, whereas E2 only annotated sleep spindles in 15 recordings. Furthermore, E1 used standard guidelines whereas E2 had access to the sigma filtered EEG signal and did not use a minimum duration, departing from standard guidelines. 

Because the filtered signal allowed E2 to see spindle activity that is obscured in the original signal, E2 annotated twice as many spindles as E1 and most E1 annotations are contained in longer E2 annotations. As a consequence, using the union of experts is practically the same as using E2 alone, and using the intersection of experts is practically the same as using E1 alone. For this reason, we did not combine the expert annotations.

The 15 recordings that have annotations from both experts were used. Because E2 did not impose a minimum duration for sleep spindles, annotations shorter than 0.3~s were removed as in \cite{warby2014sleep}. N2 epochs from the C3-CLE EEG channel were used because annotated sleep events are available only for this sleep stage and channel. On average, each recording has $720\pm 93$ N2 epochs ($240.02\pm 30.91$~min), $663\pm 318$ sleep spindles according to E1 ($9.35\pm 4.67$~min), $1446\pm 467$ sleep spindles according to E2 ($29.41\pm 10.99$~min), and $575\pm 227$ K-complexes ($7.08\pm 2.98$~min).

We selected a representative testing set consisting of the recordings 2, 6, 12, and 13, leaving 11 recordings for training and validation (see the Appendix for details). Testing recordings are fixed throughout all experiments. Therefore, the evaluation is made using data from subjects not seen during training and validation.

\subsection{Evaluation Framework}
\label{sec:eval}

Cross-validation with 10 different splits of the 11 non-testing recordings is used for evaluation purposes. The splits are generated by randomly selecting 3 validation and 8 training recordings, and they are fixed throughout all experiments. After training, we measure the test performance and report the average performance across splits. To obtain the performance of a subset, we average the performance obtained in each recording so that they are equally important. To assess the significance of the results, we use a Welch's t-test \cite{welch1947generalization} with a significance level of $0.05$. Given the annotations available in MASS, we use three tasks: sleep spindle detection according to expert E1 (SS-E1), sleep spindle detection according to expert E2 (SS-E2), and K-complex detection (KC).

For the evaluation metrics, we follow the by-event analysis of \cite{warby2014sleep} where a True Positive (TP) is credited using the Intersection over Union (IoU) between a real event $A$ and a prediction $B$, defined as
\begin{equation}
    \mathrm{IoU}(A,B)=\frac{|A\cap B|}{|A\cup B|},
    \label{eq:iou}
\end{equation}
with an ideal value of 1. First, a matching that maximizes the IoU is found between expert annotations and detections. Each of the paired events constitutes a TP-candidate. Next, IoU of matchings above a defined threshold are labeled as TPs. The remaining real events are labeled as False Negatives (FN) and the remaining detections are labeled as False Positives (FP). Unmatched real events and predictions are by definition FN and FP, respectively. Using this framework, we report the following metrics:
\begin{align}
\text{Recall} &= \frac{\text{TP}}{\text{TP}+\text{FN}}\label{eq:recall},\\
\text{Precision} &= \frac{\text{TP}}{\text{TP}+\text{FP}}\label{eq:precision},\\
\text{F1-score} &= 2\left(\frac{1}{\text{Recall}}+\frac{1}{\text{Precision}}\right)^{-1}.\label{eq:f1_score}
\end{align}

Metrics with an IoU threshold of 0.2 are reported as in \cite{warby2014sleep}, but we also analyze the robustness of the metrics by computing an F1-score versus IoU threshold curve. We propose to measure the area under this curve, here called Average F1-score (AF1), to summarize the performance at different levels of agreement. With a perfect match, the AF1 is maximum and equal to 1.
All hyperparameters and design choices for RED-Time and RED-CWT are chosen to maximize the average validation AF1 across the three tasks. The same model configuration is fixed for all tasks. The selected hyperparameters are shown in Sec. \ref{sec:result_hyper}. An important exception is the output threshold $\mu$, which is adjusted after training with a grid search between 0 and 1 with a step of 0.02 to maximize the AF1 on the non-testing set. Experimentally we found this procedure to be better than using only the validation set.

\subsection{Selected Baselines}
\label{sec:baselines}

For a fair comparison, we selected strong baselines with available open-source implementations to assess them using the same evaluation setting and the same postprocessing described in the previous sections. For sleep spindle detection, we use DOSED \cite{chambon2019dosed} and A7 \cite{lacourse2019sleep}, where the latter is a signal processing method specific to sleep spindles. SpindleNet \cite{kulkarni2019deep} was not evaluated in this common framework because it is not open-source. For completeness, we add the reported results for the union of experts in the task SS-E2 (see Sec. \ref{sec:data_description} for an explanation), but they are not directly comparable. For K-complex detection, we use DOSED and Spinky \cite{lajnef2017meet}, where the latter is a signal processing method to detect both K-complexes and sleep spindles. We did not use Spinky for sleep spindle detection because its performance in MASS is poor. Spinky provides only the location of the negative peak of the K-complex so we assume that it starts 0.1~s before and ends 1.3~s after this peak as in \cite{lajnef2017meet}. We adjusted the single threshold of Spinky and the four thresholds of A7 with a grid search to maximize the AF1 in the training set. The output threshold of DOSED was adjusted after training using the same procedure used in our models because it produces better results.

\section{Results}
In this section, we evaluate RED-Time and RED-CWT, our proposed deep learning models to detect sleep EEG events from EEG segments. We select the best combination of hyperparameters and then fix them for all experiments, and evaluate three detection tasks: sleep spindle detection according to expert E1 (SS-E1), sleep spindle detection according to expert E2 (SS-E2), and K-complex detection (KC).

\subsection{Selected Hyperparameters}
\label{sec:result_hyper}

\begin{table}[tbp]
\centering
\caption{Optimized hyperparameters for our models.}
\label{tab:hyper}
\begin{tabular}{ll}
    \toprule
    Hyperparameter  & Value \\
    \midrule
    Initial wavelet width $\beta$ & 0.5\\
    Number of scales $N_s$ & 32\\
    Initial number of filters $N_f$ (RED-CWT) & 32\\
    Initial number of filters $N_f$ (RED-Time) & 64\\
    Type of pooling & avg-pool\\
    LSTM size $N_1$ & 256 \\
    Classifier hidden size $N_2$ & 128 \\
    Dropout rate $\rho_1$ & 0.2 \\
    Dropout rate $\rho_2$ & 0.5 \\
    Batch size $M$ & 32\\
    Initial learning rate $\alpha$ & 0.0001\\
    Learning rate halving patience $I_{lr}$ & 1000\\
    \bottomrule
\end{tabular}

\end{table}

The best combination of hyperparameters were found by maximizing the average AF1 in the validation set (see Sec. \ref{sec:eval}), considering the overall performance on the three detection tasks. In Table \ref{tab:hyper} we summarize the values obtained for the hyperparameters. All experiments use the same configuration for our models. When using this configuration on a single GPU (NVIDIA GeForce GTX 1080 Ti), RED-Time and RED-CWT spend $208\pm 5$ and $352\pm 7$ seconds per 1000 iterations, respectively. Training stops at $7700\pm 900$ iterations for both models.

\subsection{Detection Performance}
\label{sec:result_performance}

\begin{table}[tbp]
\centering
\begin{threeparttable}
\caption{Detection performance of the proposed models and the baselines using an IoU threshold of 0.2.}
\label{tab:performance}
\begin{tabular}{lllll}
    \toprule
    Task & Method & F1-score & Recall & Precision \\
    \midrule
    SS-E1 & \textbf{RED-CWT} &
    $80.9\pm0.4$ & $81.5\pm1.6$ & $81.2\pm1.2$\\
    & \textbf{RED-Time} &
    $81.2\pm0.5$ \tnote{b} & $81.7\pm2.2$ & $81.8\pm1.8$\\
    & DOSED \cite{chambon2019dosed} &
    $78.4\pm0.8$ \tnote{c} & $78.3\pm2.0$ & $80.0\pm1.9$ \\
    & A7 \cite{lacourse2019sleep} &
    $71.4\pm0.3$ \tnote{c} & $75.3\pm0.4$ & $69.3\pm0.8$ \\
    \midrule
    SS-E2 & \textbf{RED-CWT} &
    $84.7\pm0.4$ & $82.6\pm1.2$ & $88.1\pm1.1$\\
    & \textbf{RED-Time} &
    $84.5\pm0.3$ \tnote{b} & $82.9\pm1.0$ & $87.4\pm1.1$\\
    & DOSED \cite{chambon2019dosed} &
    $82.0\pm0.6$ \tnote{c} & $79.2\pm1.3$ & $87.0\pm0.7$\\
    & A7 \cite{lacourse2019sleep} &
    $73.9\pm0.4$ \tnote{c} & $82.5\pm2.0$ & $68.0\pm1.0$\\
    & SpindleNet \cite{kulkarni2019deep} \tnote{a} &
    $83.0\pm 2.0$& $85.2$\hspace{21pt} & $81.0\pm 3.2$ \\
    \midrule
    KC & \textbf{RED-CWT} &
    $82.8\pm0.4$ & $81.7\pm1.1$ & $84.9\pm0.4$\\
    & \textbf{RED-Time} &
    $82.6\pm0.4$ \tnote{b} & $82.1\pm1.3$ & $83.9\pm0.8$\\
    & DOSED \cite{chambon2019dosed} &
    $77.1\pm0.8$ \tnote{c} & $76.5\pm1.2$ & $78.6\pm1.8$\\
    & Spinky \cite{lajnef2017meet} &
    $64.9\pm0.1$ \tnote{c} & $62.6\pm0.5$ & $68.7\pm0.6$\\
    \bottomrule
\end{tabular}

\begin{tablenotes}
\footnotesize
\item[a]{As reported in \cite{kulkarni2019deep}. Recall was computed from the mean values of the reported F1-score and precision. See Sec.~\ref{sec:baselines} for details.}
\item[b]{$P=0.15$ for SS-E1, $P=0.35$ for SS-E2, and $P=0.20$ for KC, compared to RED-CWT using Welch's t-test (not significant).}
\item[c]{$P<0.001$ compared to RED-CWT using Welch's t-test (significant).}
\end{tablenotes}
\end{threeparttable}
\end{table}

\begin{figure*}
\centering 
\includegraphics[width=0.85\textwidth, trim={0.09in 0.49in 0.76in 0.5in}, clip]{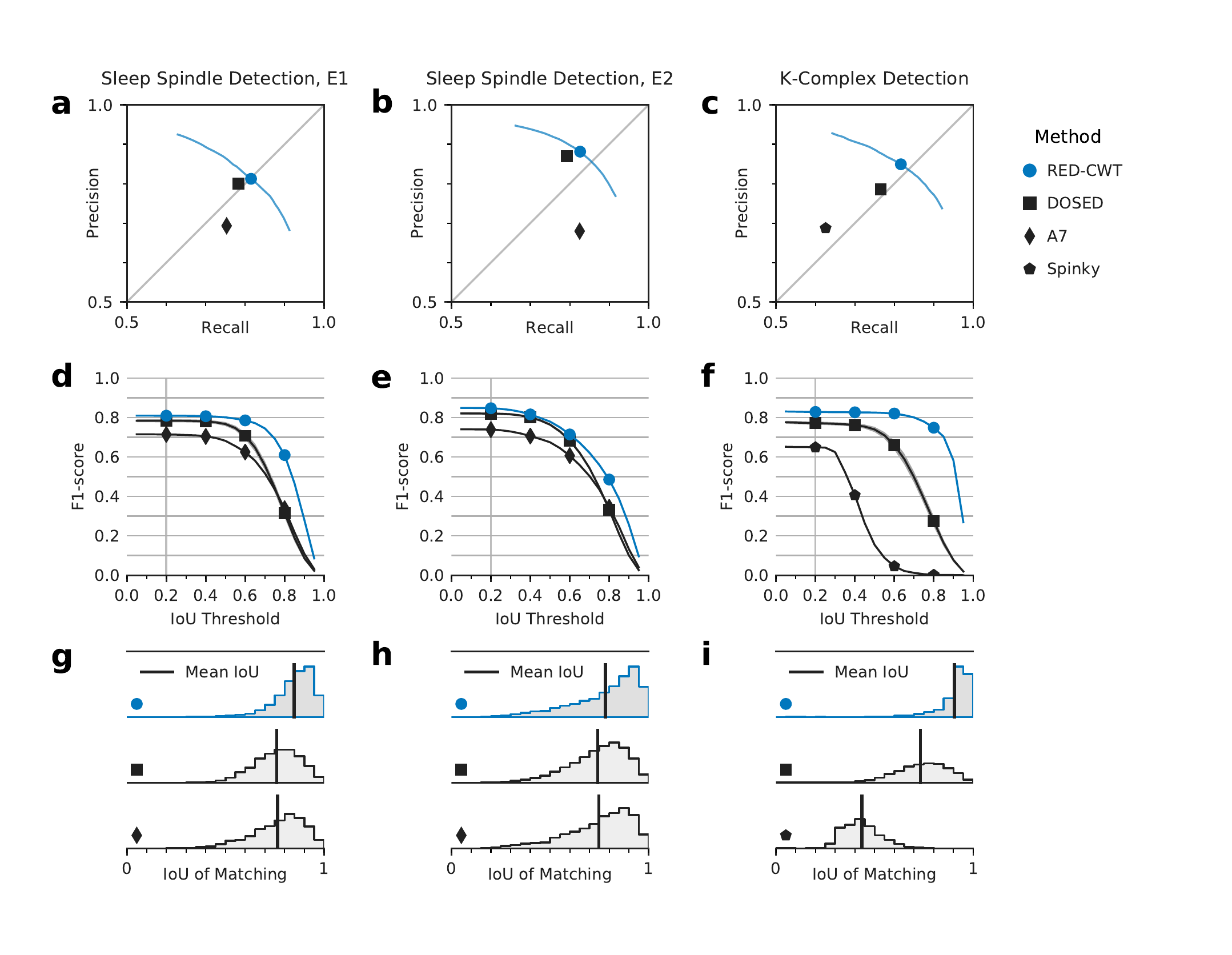}
\caption{
RED-CWT outperforms baselines in the tasks of sleep spindle detection according to expert E1 (first column), sleep spindle detection according to expert E2 (second column), and K-complex detection (third column). (\textbf{a})-(\textbf{c}) Recall and precision when $\text{IoU}\ge 0.2$. In addition, a curve was obtained for our model by varying its output threshold $\mu$. (\textbf{d})-(\textbf{f}) F1-score as a function of the IoU threshold. (\textbf{g})-(\textbf{i}) Histograms of the IoU of matchings (TP-candidates).}
\label{fig:performance}
\end{figure*}

Table \ref{tab:performance} summarizes the F1-score, recall, and precision for all models and tasks. RED-CWT and RED-Time outperform baselines on all the tasks as measured by the F1-scores, but they show no significant differences between them. The F1-score is increased by at least 2.5\% for sleep spindle detection and up to 5.7\% for K-complex detection compared with DOSED. Not only the F1-score is higher, but the gap between precision and recall for our models is generally small, which means a more balanced trade-off between FPs and FNs. Compared with the reported result of SpindleNet, the F1-score is increased by 1.7\% for SS-E2. Subsequent analysis is focused on RED-CWT, but similar findings apply to RED-Time.

The operating point in Table~\ref{tab:performance} offers the maximum F1-score. Figs.~\ref{fig:performance}a-c show precision-recall curves generated by evaluating a range of output thresholds $\mu$ from 0.1 to 0.9. For all tasks, the proposed model Pareto-dominates the operating points for the baselines shown in Table~\ref{tab:performance}.

To observe how the results shown in Table~\ref{tab:performance} are affected by the IoU threshold, we show the F1-score as a function of this threshold in Figs.~\ref{fig:performance}d-f, and the histogram of the IoU values of all matchings or TP-candidates in Figs.~\ref{fig:performance}g-i. In general, the F1-score is mostly unchanged until a threshold of 0.2, which is explained by the negligible number of TP-candidates that are discarded when using this threshold. Therefore, the results shown in Table~\ref{tab:performance} are mostly insensitive to intersection agreement. This dimension of performance is analyzed using the results shown in Figs.~\ref{fig:performance}d-i. The proposed model outperforms baselines for all IoU thresholds, obtaining a higher AF1. Moreover, the performance of our model is more robust to the IoU threshold because it decreases more slowly than the baselines as shown in Figs.~\ref{fig:performance}d-f. This is better explained when the histograms in Figs.~\ref{fig:performance}g-i are observed because our model has higher IoU values than baselines on average. 

The intersection agreement is better in the proposed models for all tasks. In sleep spindle detection, our models and the baselines have a very small portion of IoU values below 0.5. However, our models has a higher mean IoU for TP-candidates. The IoU values are more spread when using the expert E2 instead of E1 for all models. In this case, the increase in intersection agreement is still significant but smaller than in the case of using E1. The intersection agreement of our models is remarkably better in K-complex detection. It achieves a mean IoU of 0.9, and TP-candidates are mostly unaffected up to a threshold of 0.7. On the other hand, DOSED has a mean IoU slightly greater than 0.7 with a larger dispersion. In this task, low intersection agreement is obtained with Spinky, which is expected due to its fixed duration assumption.

For K-complex detection, the context needed around the event to accurately determine the start and end times is larger than for sleep spindle detection, where a recognition of presence or absence of oscillatory activity might be enough. The difference in performance is notably large in K-complex detection even when DOSED processes segments of equal length, which suggests that RED makes a better use of context.

\begin{figure}[tbp]
\centering 
\includegraphics[width=0.85\columnwidth, trim={0.13in 0.17in 0.17in 0.16in}, clip]{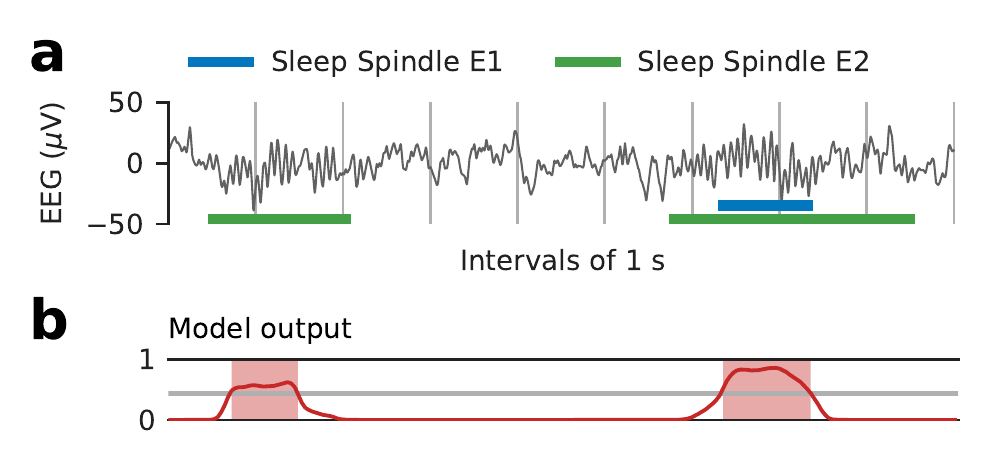}
\caption{
\textbf{(a)} A testing EEG segment with sleep spindle annotations from experts E1 and E2. \textbf{(b)} Model output (solid red line), output threshold (solid gray line) and final predictions (shaded rectangles) after training on E1's annotations. The figure shows an FP and a TP case, where the former matches partially an E2's annotation.
}
\label{fig:example_e1_e2_pred}
\end{figure}

We can further analyze the false positives in sleep spindle detection using the two available experts. After training with E1's annotations and assigning TPs using $\text{IoU}>0$, more than 90\% of FPs have positive intersection with E2's annotations. Hence, most FPs with respect to E1 criterion are at least reasonable predictions according to E2. However, when E1 and E2 agree on the existence of a sleep spindle, E2 annotated a longer event in general, resulting in a mean IoU of 0.6 between experts. Therefore, a low IoU between the model and E2 is expected as well because the model learned E1's preferences regarding the duration of the events (see Fig.~\ref{fig:example_e1_e2_pred}).

\section{Conclusions}

We have proposed RED, a deep learning approach for sleep EEG event detection. Thanks to the recurrent layers applied across the entire EEG segment input, the method has more capacity to model long-term temporal context than those of previous works using deep learning methods for detecting sleep-spindles and K-complexes. As hypothesized, this capability improved the detection performance, measured as intersection agreement. By predicting a dense segmentation, RED does not define an arbitrary partition in the input segment. Moreover, it does not define a context window because the recurrent layers learn how much context is needed at each time step. These characteristics make the approach simpler so that it is more likely to generalize to different sleep EEG events. Using the same model, we achieved state-of-the-art performance both in sleep spindle and K-complex detection.

A drawback of deep learning models is the lack of explanation as a result of behaving like black boxes. Conversely, the predictions of traditional methods can be explained by inspecting their features. This has been widely recognized as an important factor in areas like medicine \cite{adadi2018peeking}. The results obtained show that RED-CWT and RED-Time have a similar performance, but we conjecture that RED-CWT might offer better interpretability because relevant inputs at different time instants and different frequency bands could be highlighted in the spectrogram. Testing this conjecture is left for future work.

The deep learning model can learn the specific expert bias contained in the annotated dataset. Therefore, the model could be made more useful in the clinical practice by adopting one of two paths. The first one is to train on a high-quality annotated dataset using the consensus of a group of experts as in \cite{warby2014sleep} so that the learned model predicts in a standardized way. The second one is to adapt a trained model to each expert's preferences. This problem is left as future work.

\appendix

We partitioned the set of 15 recordings from MASS into 4 testing recordings and 11 non-testing recordings, so that the testing set is representative of the non-testing set. To achieve this, we visually checked the projection of the FFT of the EEG signals. First, each N2 epoch of the C3-CLE EEG channel is collected from all recordings. For each epoch $x_i$, its FFT is computed. Five frequency bands $B_j$ with frequencies $f$ between $f^L_j$ and $f^U_j$ are determined based on bands used in sleep medicine \cite{berry2012aasm}, leading to
\begin{equation}
(f^L_j, f^U_j) \in \{(1,4), (4,8), (8,12), (12,15), (15,30)\}~\text{Hz}.
\label{eq:freq_band_limits}
\end{equation}
Next, the average power within each band is computed as
\begin{equation}
    \bar{b}_{ij} = \log\left(\frac{1}{|B_j|}\sum_{f\in B_j} |\text{FFT}[x_i](f)| \right).
    \label{eq:average_power}
\end{equation}

\begin{figure}[tbp]
\centering 
  \includegraphics[width=0.8\columnwidth, trim={0.17in 0.18in 0.16in 0.08in}, clip]{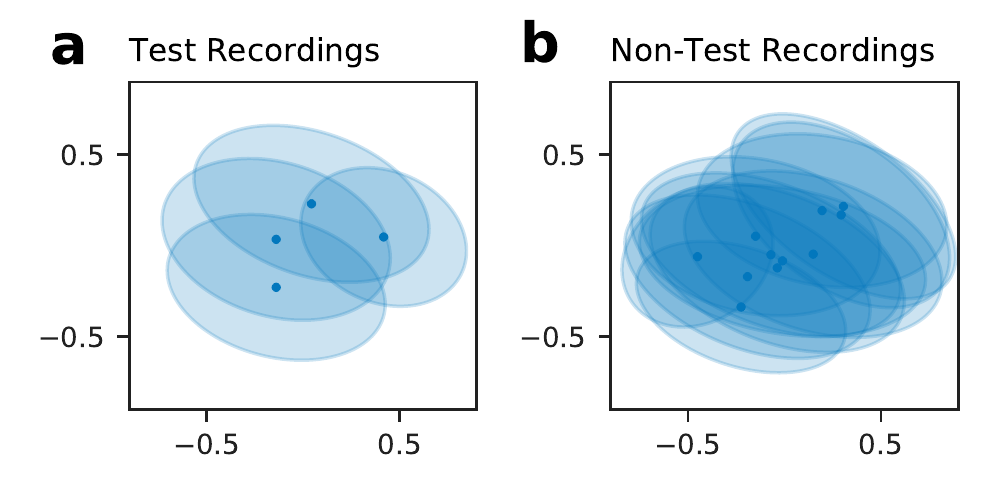}
  \caption{Kernel PCA projections of the testing and non-testing recordings based on power features. (\textbf{a}) Fitted Gaussian distributions of testing recordings 2, 6, 12, and 13. (\textbf{b}) Fitted Gaussian distributions of non-testing recordings.}
  \label{fig:data_split}
\end{figure}

These features are standardized and projected in 2D using Kernel PCA \cite{scholkopf1997kernel} with RBF kernel. The kernel coefficient $\gamma$ is chosen from $\gamma\in\{0.01, 0.1, 1, 10\}$ for a good visualization, leading to $\gamma=0.1$. In this projected space, a 2D Gaussian distribution is fitted to each set of epochs belonging to the same recording. Next, we randomly generate testing sets and visually judge whether the combined distribution of the testing set is representative of the non-testing set. We achieved a representative testing set with the recordings 2, 6, 12, and 13 as illustrated in Fig.~\ref{fig:data_split}, where the Gaussian distributions are shown with their mean and their 95\% confidence ellipses.

\bibliographystyle{IEEEtran}
\bibliography{IEEEabrv, references}

\begin{thebibliography}{10}
\providecommand{\url}[1]{#1}
\csname url@samestyle\endcsname
\providecommand{\newblock}{\relax}
\providecommand{\bibinfo}[2]{#2}
\providecommand{\BIBentrySTDinterwordspacing}{\spaceskip=0pt\relax}
\providecommand{\BIBentryALTinterwordstretchfactor}{4}
\providecommand{\BIBentryALTinterwordspacing}{\spaceskip=\fontdimen2\font plus
\BIBentryALTinterwordstretchfactor\fontdimen3\font minus
  \fontdimen4\font\relax}
\providecommand{\BIBforeignlanguage}[2]{{%
\expandafter\ifx\csname l@#1\endcsname\relax
\typeout{** WARNING: IEEEtran.bst: No hyphenation pattern has been}%
\typeout{** loaded for the language `#1'. Using the pattern for}%
\typeout{** the default language instead.}%
\else
\language=\csname l@#1\endcsname
\fi
#2}}
\providecommand{\BIBdecl}{\relax}
\BIBdecl

\bibitem{berry2012aasm}
R.~Berry, R.~Brooks, C.~Gamaldo, S.~Harding, R.~Lloyd, C.~Marcus, and
  B.~Vaughn, \emph{The {AASM} Manual for the Scoring of Sleep and Associated
  Events: Rules, Terminology and Technical Specifications, Version 2.5}.\hskip
  1em plus 0.5em minus 0.4em\relax American Academy of Sleep Medicine, 2018.

\bibitem{coppieters2016sleep}
D.~Coppieters, P.~Maquet, and C.~Phillips, ``Sleep spindles as an
  electrographic element: Description and automatic detection methods,''
  \emph{Neural Plast.}, vol. 2016, p. 6783812, 2016.

\bibitem{clawson2016form}
B.~Clawson, J.~Durkin, and S.~Aton, ``Form and function of sleep spindles
  across the lifespan,'' \emph{Neural Plast.}, vol. 2016, p. 6936381, 2016.

\bibitem{wauquier1995k}
A.~Wauquier, L.~Aloe, and A.~Declerck, ``K-complexes: Are they signs of arousal
  or sleep protective?'' \emph{J. Sleep Res.}, vol.~4, no.~3, pp. 138--143,
  1995.

\bibitem{el2008k}
J.~El~Helou, V.~Navarro, C.~Depienne, E.~Fedirko, E.~LeGuern, M.~Baulac,
  I.~An-Gourfinkel, and C.~Adam, ``K-complex-induced seizures in autosomal
  dominant nocturnal frontal lobe epilepsy,'' \emph{Clin. Neurophysiol.}, vol.
  119, no.~10, pp. 2201--2204, 2008.

\bibitem{guilleminault1976sleep}
C.~Guilleminault, A.~Tilkian, and W.~Dement, ``The sleep apnea syndromes,''
  \emph{Annu. Rev. Med.}, vol.~27, no.~1, pp. 465--484, 1976.

\bibitem{glasauer2001restless}
F.~Glasauer, ``Restless legs syndrome,'' \emph{Spinal Cord}, vol.~39, no.~3,
  pp. 125--133, 2001.

\bibitem{warby2014sleep}
S.~Warby, S.~Wendt, P.~Welinder, E.~Munk, O.~Carrillo, H.~Sorensen, P.~Jennum,
  P.~Peppard, P.~Perona, and E.~Mignot, ``Sleep-spindle detection:
  Crowdsourcing and evaluating performance of experts, non-experts and
  automated methods,'' \emph{Nat. Methods}, vol.~11, no.~4, pp. 385--392, 2014.

\bibitem{chambon2019dosed}
S.~Chambon, V.~Thorey, P.~Arnal, E.~Mignot, and A.~Gramfort, ``{DOSED}: A deep
  learning approach to detect multiple sleep micro-events in {EEG} signal,''
  \emph{J. Neurosci. Methods}, vol. 321, pp. 64--78, 2019.

\bibitem{kulkarni2019deep}
P.~Kulkarni, Z.~Xiao, E.~Robinson, A.~Jami, J.~Zhang, H.~Zhou, S.~Henin,
  A.~Liu, R.~Osorio, J.~Wang, and Z.~Chen, ``A deep learning approach for
  real-time detection of sleep spindles,'' \emph{J. Neural Eng.}, vol.~16,
  no.~3, p. 036004, 2019.

\bibitem{addison2017illustrated}
P.~Addison, \emph{The illustrated wavelet transform handbook: Introductory
  theory and applications in science, engineering, medicine and finance}.\hskip
  1em plus 0.5em minus 0.4em\relax CRC press, 2017.

\bibitem{parekh2017multichannel}
A.~Parekh, I.~Selesnick, R.~Osorio, A.~Varga, D.~Rapoport, and I.~Ayappa,
  ``Multichannel sleep spindle detection using sparse low-rank optimization,''
  \emph{J. Neurosci. Methods}, vol. 288, pp. 1--16, 2017.

\bibitem{larocco2018spindler}
J.~LaRocco, P.~Franaszczuk, S.~Kerick, and K.~Robbins, ``Spindler: A framework
  for parametric analysis and detection of spindles in {EEG} with application
  to sleep spindles,'' \emph{J. Neural Eng.}, vol.~15, no.~6, p. 066015, 2018.

\bibitem{held2004dual}
C.~Held, L.~Causa, P.~Est{\'e}vez, C.~P{\'e}rez, M.~Garrido, C.~Algar{\'\i}n,
  and P.~Peirano, ``Dual approach for automated sleep spindles detection within
  {EEG} background activity in infant polysomnograms,'' in \emph{Proc. 26th
  Annu. Int. Conf. IEEE Eng. Med. Biol. Soc.}, 2004, pp. 566--569.

\bibitem{lacourse2019sleep}
K.~Lacourse, J.~Delfrate, J.~Beaudry, P.~Peppard, and S.~Warby, ``A sleep
  spindle detection algorithm that emulates human expert spindle scoring,''
  \emph{J. Neurosci. Methods}, vol. 316, pp. 3--11, 2019.

\bibitem{lachner2018single}
D.~Lachner-Piza, N.~Epitashvili, A.~Schulze-Bonhage, T.~Stieglitz, J.~Jacobs,
  and M.~D{\"u}mpelmann, ``A single channel sleep-spindle detector based on
  multivariate classification of {EEG} epochs: {MUSSDET},'' \emph{J. Neurosci.
  Methods}, vol. 297, pp. 31--43, 2018.

\bibitem{lajnef2017meet}
T.~Lajnef, C.~O’Reilly, E.~Combrisson, S.~Chaibi, J.-B. Eichenlaub, P.~Ruby,
  P.-E. Aguera, M.~Samet, A.~Kachouri, S.~Frenette, J.~Carrier, and K.~Jerbi,
  ``Meet {Spinky}: an open-source spindle and {K}-complex detection toolbox
  validated on the open-access {M}ontreal {A}rchive of {S}leep {S}tudies
  ({MASS}),'' \emph{Front. Neuroinform.}, vol.~11, p.~15, 2017.

\bibitem{estevez2007sleep}
P.~Est{\'e}vez, R.~Zilleruelo-Ramos, R.~Hern{\'a}ndez, L.~Causa, and C.~Held,
  ``Sleep spindle detection by using merge neural gas,'' in \emph{Proc. 6th
  Int. Workshop Self-Organizing Maps (WSOM)}, 2007.

\bibitem{causa2010automated}
L.~Causa, C.~Held, J.~Causa, P.~Est{\'e}vez, C.~Perez, R.~Chamorro, M.~Garrido,
  C.~Algar{\'\i}n, and P.~Peirano, ``Automated sleep-spindle detection in
  healthy children polysomnograms,'' \emph{{IEEE} Trans. Biomed. Eng.},
  vol.~57, no.~9, pp. 2135--2146, 2010.

\bibitem{ulloa2016sleep}
S.~Ulloa, P.~Est{\'e}vez, P.~Huijse, C.~Held, C.~Perez, R.~Chamorro,
  M.~Garrido, C.~Algar{\'\i}n, and P.~Peirano, ``Sleep-spindle identification
  on {EEG} signals from polysomnographie recordings using correntropy,'' in
  \emph{Proc. 38th Annu. Int. Conf. IEEE Eng. Med. Biol. Soc.}, 2016, pp.
  3736--3739.

\bibitem{luo2016understanding}
W.~Luo, Y.~Li, R.~Urtasun, and R.~Zemel, ``Understanding the effective
  receptive field in deep convolutional neural networks,'' in \emph{Advances in
  Neural Information Processing Systems 29}, 2016, pp. 4898--4906.

\bibitem{ioffe2015batch}
S.~Ioffe and C.~Szegedy, ``Batch normalization: Accelerating deep network
  training by reducing internal covariate shift,'' in \emph{Proc. 32nd Int.
  Conf. Machine Learning ({ICML})}, 2015, pp. 448--456.

\bibitem{hochreiter1997long}
S.~Hochreiter and J.~Schmidhuber, ``Long short-term memory,'' \emph{Neural
  Comput.}, vol.~9, no.~8, pp. 1735--1780, 1997.

\bibitem{graves2005framewise}
A.~Graves and J.~Schmidhuber, ``Framewise phoneme classification with
  bidirectional {LSTM} and other neural network architectures,'' \emph{Neural
  Networks}, vol.~18, no. 5-6, pp. 602--610, 2005.

\bibitem{srivastava2014dropout}
N.~Srivastava, G.~Hinton, A.~Krizhevsky, I.~Sutskever, and R.~Salakhutdinov,
  ``Dropout: A simple way to prevent neural networks from overfitting,''
  \emph{J. Mach. Learn. Res.}, vol.~15, no.~1, pp. 1929--1958, 2014.

\bibitem{kingma2014adam}
D.~Kingma and J.~Ba, ``Adam: {A} method for stochastic optimization,'' in
  \emph{Int. Conf. Learning Representations ({ICLR})}, 2015.

\bibitem{pascanu2013difficulty}
R.~Pascanu, T.~Mikolov, and Y.~Bengio, ``On the difficulty of training
  recurrent neural networks,'' in \emph{Proc. 30th Int. Conf. Machine Learning
  ({ICML})}, 2013, pp. 1310--1318.

\bibitem{o2014montreal}
C.~O'Reilly, N.~Gosselin, J.~Carrier, and T.~Nielsen, ``Montreal {A}rchive of
  {S}leep {S}tudies: an open-access resource for instrument benchmarking and
  exploratory research,'' \emph{J. Sleep Res.}, vol.~23, no.~6, pp. 628--635,
  2014.

\bibitem{welch1947generalization}
B.~Welch, ``The generalization of '{S}tudent's' problem when several different
  population variances are involved,'' \emph{Biometrika}, vol.~34, no. 1/2, pp.
  28--35, 1947.

\bibitem{adadi2018peeking}
A.~Adadi and M.~Berrada, ``Peeking inside the black-box: A survey on
  explainable artificial intelligence ({XAI}),'' \emph{IEEE Access}, vol.~6,
  pp. 52\,138--52\,160, 2018.

\bibitem{scholkopf1997kernel}
B.~Sch{\"o}lkopf, A.~Smola, and K.-R. M{\"u}ller, ``Kernel principal component
  analysis,'' in \emph{Int. Conf. Artificial Neural Networks (ICANN)}, 1997,
  pp. 583--588.

\end{thebibliography}
\end{document}